\documentclass[preprint,showpacs,showkeys,preprintnumbers,amsmath,amssymb]{revtex4}

\usepackage[pdftex]{graphicx}
\usepackage{bm}
\usepackage{color}
\usepackage{mathrsfs}

\begin{document}

\texttt{preprint}
\preprint{APS/123-QED}

\title{Phase Slip Avalanches in Small Superconductors}

\author{M.B. Sobnack}
\author{F.V. Kusmartsev}
\affiliation{
Department of Physics, Loughborough University, Loughborough LE11 3TU,
United Kingdom}
\date{\today}

\begin{abstract}

We study the effect of phase slips in a quasi 1d superconducting channel along which a current flows and report a new phenomenon where an avalanche of phase slips occurs. This limits the critical current in thin films and wires and drives the system to a topological phase transition at a temperature lower than the bulk critical temperature. We describe the mechanism of such a catastrophic phase slip avalanche and, following Kosterlitz and Thouless, we use group renormalization techniques to derive an exact analytical expression for the critical current as a function of film width and temperature. Our results are in very good agreement with, and reproduce, the available experimental data on superconducting MgB$_2$ thin films. The phenomenon we describe is very general and can be used in the construction of new devices where the superconducting state can coexist with the normal state. 
\end{abstract}


\maketitle

It is well known that phase slips give rise to the appearance of resistance in a thin superconducting wire below the bulk transition temperature $T_c$ \cite{Tinkham,Arutyunov1}. During a phase slip event, the superconducting order parameter vanishes at some point along the width of the wire and the phase slips by 2$\pi$ there; this is accompanied by a voltage drop pulse and, if many such events take place, a resistive voltage will appear in the wire, {\it i.e.}, it will lead to dissipative behaviour. Phase slip has been the subject of many studies in recent years. It has been widely studied in ultra narrow (width $\leqslant \xi$, the coherence length) \cite{Zgirski,Elmurodov,PengLi,Fomin,Croitaru-2012A,Arutyunov-PRL-2012,Cirillo-2012,Croitaru-2012B,Rogachev-2012,Nazarov-2012,Arutyunov-SR-2012} and to a lesser extent in wide (width $>\xi$) superconducting wires/strips \cite{Sivakov,Tettamanzi}. It has been shown that the character of the superconducting phase slip depends on the transverse
size of the superconducting nanowires and that for ultra-narrow nanowires, the phase slip ascribes
a quantum character that changes the character of the superconducting state and can even
completely suppress it \cite{Zgirski}.

\begin{figure}
  \centering
  \includegraphics[width=15cm]{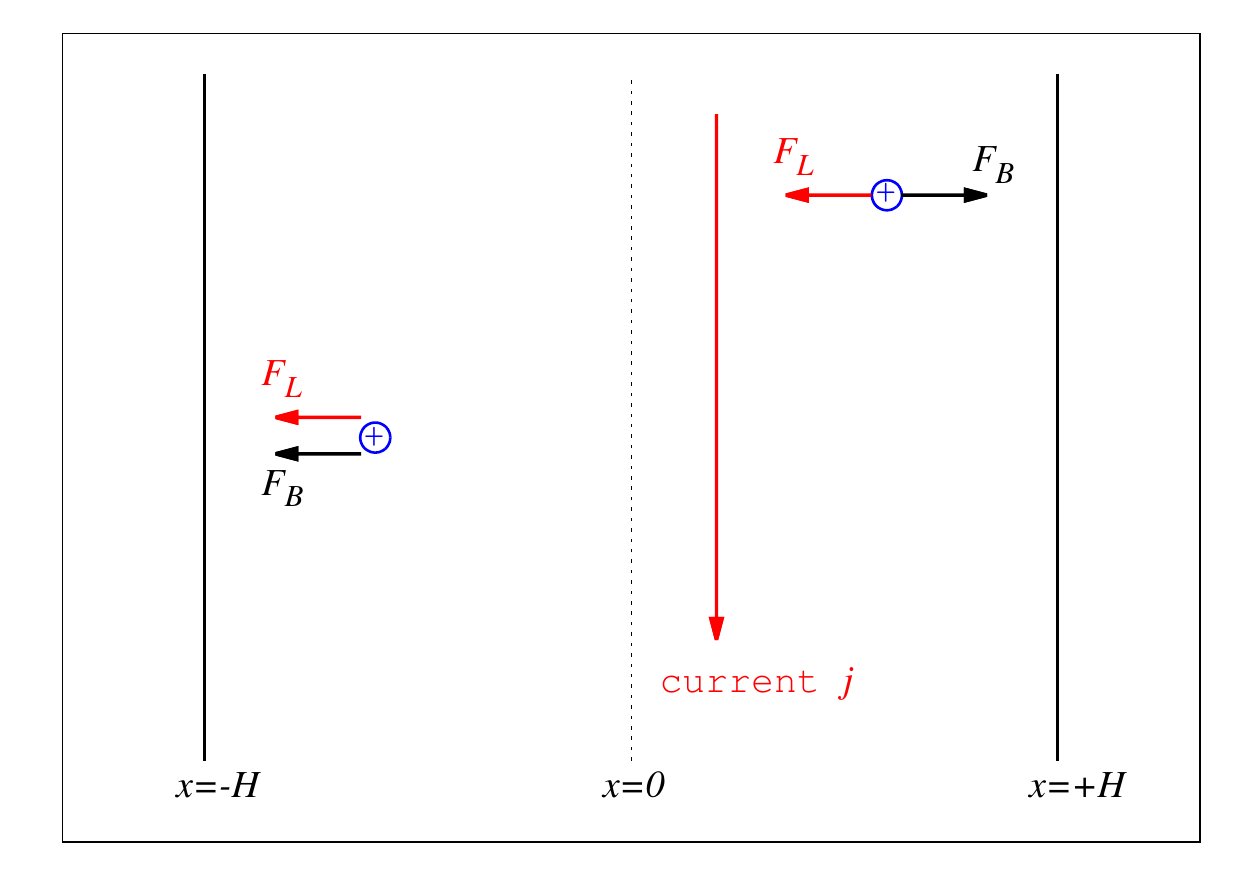}
  \caption{The figure shows schematically the superconducting channel (of width $2H$) studied in the paper.  A vortex is strongly attracted to the boundary by a force $\mathbf{F_{\!B}}$ and the energy required to create such a vortex is obtained by calculating the energy of interaction of the vortex with an infinite number of image vortices and an infinite number of image anti-vortices situated beyond the channel. In the presence of a current $\mathbf{j}=-j\mathbf{\hat{y}}$ in the channel, the vortex also experiences a Lorentz force $\mathbf{F_{\!L}}$ directed away from the boundary if the vortex is located near the right-hand boundary and towards the boundary if it is near the left-hand boundary.}\label{fig:myfigure1}
\end{figure}

In this manuscript we study the suppression of  superconductivity in narrow superconducting thin tapes and nanowires and, in general, in narrow superfluid channels. In contrast to other studies \cite{Zgirski,Elmurodov,PengLi,Fomin,Croitaru-2012A,Arutyunov-PRL-2012,Cirillo-2012,Croitaru-2012B,Rogachev-2012,Nazarov-2012,Arutyunov-SR-2012}, we do not study ultra narrow superconducting wires, but superconducting/superfluid channels of widths larger than the superconducting coherence length. For narrow superconducting wires, the low energy excitations of the systems are vortices created near the boundary. We consider a mechanism of the suppression of the superconducting state which arises at high temperatures and is associated with the spontaneous cascade creation of vortices from the boundary and an avalanche of phase slips in the channels. The study is twofold: (1) in the absence of a current, we show that nucleation of single vortices in the channel drives the system to a topological phase transition at a scale-dependent temperature that is lower than the superconducting critical temperature, and (2) in the presence of a current along the channel, the suppression of superconductivity happens at an even lower temperature. 

The model we study is a quasi-1D channel of superconducting film of width $2H$, lying between $x=-H$ and $x=H$ and unbounded in the $y$-direction, with a current $\mathbf {j}$ along the length of the channel (see Fig.~1). We describe the effect of the interaction of a single vortex (or anti-vortex) with the current and with the two channel edges/boundaries. Because of the current, any fluctuational vortex in the channel will experience a Lorentz force $\mathbf{F_{\!L}}$ in addition to the attractive force $\mathbf{F_{\!B}}$ to the boundary. In the geometry shown in the figure, where the current $\mathbf{j}=-j\mathbf{\hat{y}}$ is in the negative $y$-direction, the Lorentz force $\mathbf{F_{\!L}}=-F_{\!L}\mathbf{\hat{x}}$ on a vortex near the right-hand boundary of the channel at $x=+H$ will be in opposition to the force $\mathbf{F_{\!B}}$ with which it is attracted to the boundary, whereas for one near the left-hand boundary at $x=-H$ the Lorentz force will be in the same direction as $\mathbf{F_{\!B}}$. Hence any vortex created in the left-hand half of the channel ({\it i.e.} in the region $-H<x<0$) will be very strongly attracted to the boundary at $x=-H$ and will be ``ejected" from the channel. Any vortex created in the right-hand half of the channel will be driven towards the boundary at $x=-H$ if $|\mathbf{F_{\!L}}|>|\mathbf{F_{\!B}}|$. If this happens, the phase of the order parameter will slip \cite{Tinkham} and, if there are enough of these events along the length of the channel, the channel will acquire a non-zero resistance at temperatures below the conventional transition temperature $T_c$. (For anti-vortices created in the channel, the Lorentz force will be in the opposite direction. Without loss of generality, we restrict our discussion only to vortices).

The attraction of a vortex to the boundary can be described as its attraction to infinitely many imaginary vortices (IV) and imaginary anti-vortices (IA), located beyond the sample (vortices, just like 2D electric charges, are attracted to their mirror anti-charges). A vortex can penetrate into the sample only by overcoming this attraction to its multiple images. For bulk superconductors, this attraction to the boundary is known as a surface or Bean-Livingston barrier. Let us assume that
several thermally excited vortices have been
created near the boundary. These non-stationary vortices induce infinitely many images which
renormalize (decrease) the vortex attraction to the boundary. In a
fashion similar to the screening of the vortex-antivortex interaction in the Berezinsk\u{i}-Kosterlitz-Thouless \cite{Ber,Kos1,Kos2} (BKT)
mechanism, now the imaginary plasma of V-IA, V-IV  screens
the ``Coulomb'' attraction. Eventually this chain process leads to the
nucleation of fluctuational vortices in a
relatively wide layer near the boundary and the order parameter
associated with superfluidity is destroyed. The temperature at which
this crossover happens can be defined as a new size-dependent
critical temperature which is lower than the critical temperature
$T_c$ of the bulk material. 

Consider a vortex of vorticity $\kappa=h/m$ at an arbitrary position ${\mathbf r_0}=(x_0,y_0),~0<x_0<H$, in the channel. The correct boundary condition on $x=\pm H$ is that any flow at the boundaries should be tangential, {\it i.e.}, the normal component of the flow at the boundaries should be zero. This is achieved, in analogy with 2D electrostatics, by having an infinite array of image vortices (each of vorticity $\kappa$) beyond the channel at ${\mathbf r_m}=(x_m,y_0)$ and an infinite array of image anti-vortices (of vorticity $-\kappa$) at ${\mathbf r'_m}=(x'_m,y_0)$, where
\begin{displaymath}
x_m=x_0+4Hm,~~m\in \mathbb{Z}, m\neq 0~\mathrm{and}~x'_m=2H-x_0+4Hm,~~m\in \mathbb{Z}.
\end{displaymath}
The energy required to create the vortex at ${\mathbf r_0}$ is the energy with which it is attracted to the boundaries. This can be evaluated by calculating the energy of interaction of the vortex with all its image vortices and anti-vortices. After a lengthy calculation based on classical hydrodynamics (see supplementary material), one finds that the energy required to create a vortex in the 2D channel of superfluid at $\mathbf{r_0}=(x_0,y_0)$ is 
\begin{equation}
U_0(x_0)=2q^2\ln \frac{2\cos kx_0}{kr_c}+E_c, ~~x_0\leqslant |H-r_c|,
\end{equation} 
where $r_c$ is the radius of the core of the vortex,  $k=\pi/2H$, $E_c$ is the potential energy associated with the vortex core and, where in analogy with the 2D Coulomb gas,
\begin{displaymath}
q=\frac{1}{2}\kappa \left (\frac{\rho_s}{\pi}\right )^{1/2}=\frac{\hbar}{m}\sqrt{\pi \rho_s}
\end{displaymath}
is the effective vortex charge. The core energy $E_c$
is sometimes written as $-\mu$ and $\mu$ is called the chemical potential of
the vortex. $U_0(x_0)$ is the
energy with which the vortex charge at $\mathbf{r_0}$ is attracted to the
boundary of the superfluid channel. Hencefore, we shall drop the subscript ``0" from $x_0$. 

Consider a vortex (with $\kappa=h/m$) at position $x>0$ in the channel (see Fig.~1) when there is a current $\mathbf{j}=-j\mathbf{\hat{y}}$ in the channel. The vortex experiences a Lorentz force
\begin{equation}
\mathbf{F_{\! L}}=\mathbf{j}\wedge\mathbf{\Phi_0}=-j\Phi_0\mathbf{\hat{x}},
\end{equation}
where $\mathbf{\Phi_0}=\Phi_0\mathbf{\hat{z}}$ is a vector of magnitude $\Phi_0=h/2e$, the flux quantum, directed along the vortex axis. The work done in moving the vortex from the boundary to position $x$ is
\begin{equation}
W=\int_H^x\mathbf{F_{\! L}}\cdot \mathrm{d\mathbf{x}}=\int_H^xj\Phi_0\mathrm{dx}=-j\Phi_0(H-x).
\end{equation}
Hence the total energy required to create a vortex in the channel at $x$ is
\begin{equation}
U(x)=U_0+W=2q^2\ln \frac{2\cos kx}{kr_c}-j\Phi_0(H-x)+E_c,
\end{equation}
which has a maximum value at
\begin{equation}
x_0=\frac{2H}{\pi}\tan^{-1}\frac{j\Phi_0H}{\pi q^2}\cdot
\end{equation}
In addition to the Lorentz force $\mathbf{F_{\! L}}=-j\Phi_0\mathbf{\hat{x}}$, the vortex also experiences a force $\mathbf{F_{\!B}}=-\nabla U_0(x)=2q^2k\cot kx \,\mathbf{\hat{x}}$ that attracts it to the boundary. A vortex cannot appear and start to move from the right boundary across the channel to the left boundary -- to do so it will need to overcome the potential barrier (see Fig.~2). However, if $j$ is large enough for the fluctuational vortex to reach the (unstable) equilibrium point $x_0$, a phase slip event will take place: a vortex created at the right-hand boundary ``slips" across to the left-hand boundary where it leaves the channel, accompanied by a phase change of $2\pi$. If several of these events take place along the length of the channel,  the channel will acquire a finite resistance below $T_c$ \cite{Tinkham, Langer}.

\begin{figure}
  \centering
  \includegraphics[width=15cm]{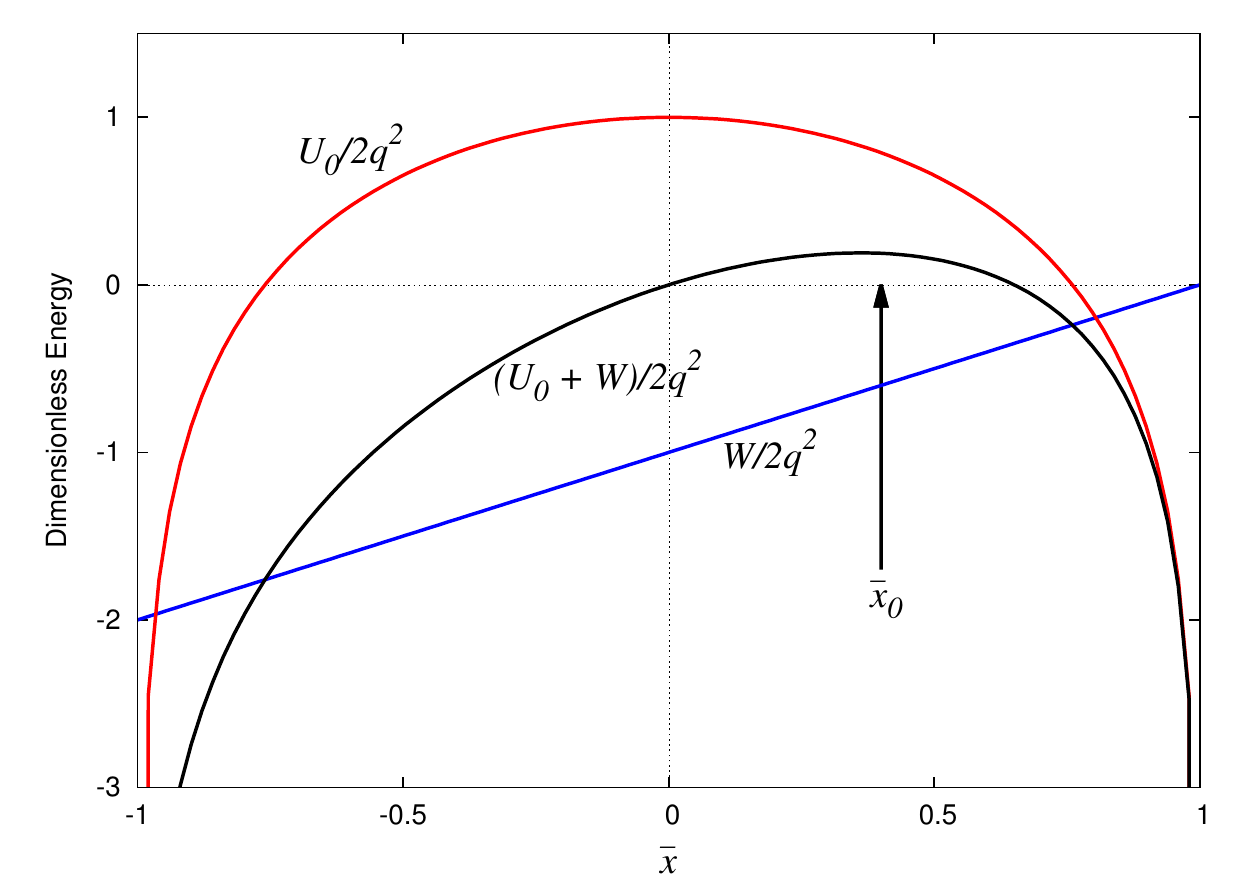}
  \caption{Plots of the dimensionless energies $U_0(\bar{x})/2q^2$ (red line), $W(\bar{x})/2q^2$ (blue line) and the total energy $[U_0(\bar{x})+W(\bar{x})]/2q^2$ (black line) as functions of $\bar{x}=x/H$. For illustration purposes, we have taken $E_c/2q^2 - \ln kr_c=1-\ln 2$ and $j\Phi_0H/2q^2=1$. $\bar{x}_0$ is the value of $\bar{x}$ at the maximum in the total energy.}\label{fig:myfigure2}
\end{figure}

At low temperatures, it is likely that only a few fluctuational vortices will be
present in the channel. These are attracted to the boundary and cannot nucleate. 
(This will also happen with virtual vortices and anti-vortices originating from a vacuum state as well.) At
higher temperatures, however, there are likely to be many more vortex
excitations, including some located in the space between $x$ and
$H$. These have an attenuating effect on, and screen, the interaction
$U_0(x)$. To take into account this screening effect, we follow
Kosterlitz and Thouless \cite{Kos1,Kos2}, Kusmartsev \cite{Kusmartsev1}, Sobnack \cite{Sob1}, Williams \cite{Wil}
and Shenoy \cite{She} by introducing a scale-dependent dielectric
constant
$
\varepsilon=1+4\pi\chi
$.
The effective susceptibility
$\chi=\int_{r_c}^{H-r_c}\alpha(x)dn(x)$, where
$\alpha(x)=q^2(H-x)^2/2k_BT$ is the polarizability and $n(x)$ is the
number density of vortices. It is straightforward to show that
$dn(x)=dx \exp (-U_{0s}(x)/k_BT)/(2\pi r_c^2H)$, where $U_{0s}(x)$ is
the screened interaction. In terms of these, and introducing the dimensionless superfluid density $K=q^2/(\pi
k_BT)$ and the renormalized density $K_r=K/\varepsilon(r)$, one arrives at 
an equation for $K_r$,
\begin{equation}
K_r^{-1}=K^{-1}+\frac{\pi \tilde{y}_0}{r_c^2H}\int_{r_c}^{H-r_c}dx\,(H-x)^2 \exp\left [-2\pi K\ln\frac{2\cos kx}{kr_c}+\frac{j\Phi_o}{k_bT}(H-x)\right ],
\end{equation}
where $\tilde{y_0}=\exp(-E_c/k_BT)$. In deriving the above equation,
we have implicitly assumed a rather low density of vortices, evident,
for example, in the fact that we have used the unrenormalized charge
$q$ instead of $q_r=q/\varepsilon(r)$ to determine the
polarizability. We have also neglected the
correction term in the interaction energy and used the bare interaction energy $U_0(x)$. These approximations
do not change the results and are necessary to prevent the equations from becoming intractable. Strictly, one needs to replace $K$ in the exponent by the
renormalized density $K_r$ to make Eq.~(6) self-consistent. However,
at low temperatures, the integral is small, and Eq.~(6) is the first
two terms in the expansion of $K_r^{-1}$.

For temperatures in the neighbourhood of the crossover where the superfluid density tends to zero, the perturbative expansion is not valid and here and we use the vortex-core rescaling technique introduced by Jos\'{e} {\it et al}. \cite{Jos}. One eventually arrives 
at the differential renormalization group (or scaling) equations  
\begin{eqnarray}
\frac{dK_l}{dl}&=&K_l-\pi\frac{(H-r_c)^2}{r_cH}K_l^2y_l\nonumber \\
& & \\
\frac{dy_l}{dl}&=&(1-2\pi K_l)y_l\nonumber
\end{eqnarray}
for the effective couplings $K_l$ and $y_l$, together with $dj_l/dl=j_l$ and $dH_l/dl=-H_l$, with the definition $dl=\ln b$. 

The critical (or fixed) point of the scaling equations is given by $dK_l/dl=0$ and $dy_l/dl=0$. These two equations have the nontrivial solution
\begin{equation}
K^*=\frac{1}{2\pi}~~\mathrm{and}~~y^*=\frac{2r_cH}{(H-r_c)^2},
\end{equation}
describing the critical point which is a saddle point. The  critical point separates the superconducting and normal states in the narrow channel. Interesting enough is the fact that the critical superconducting density does not depend on the channel width, whereas the critical value for  the vortex fugacity does depend on it.  

To investigate how the superconducting state vanishes, we analyse the scaling in the vicinity of the critical point and connect this scaling with the thermodynamic behavior of the system. We write the scaling equations (11) in terms of scaled deviations from the fixed points by expanding $K_l$ and $y_l$ around $(K^*,y^*)$ as $K_l=K^*(1+\kappa')$ and $y_l=y^*(1+y')$. To first order in $k'$ and $y'$, the scaling equations then read (we now drop the primes)
\begin{equation}
\left ( \begin{array}{c} \dot{\kappa}\\ \dot{y} \end{array} \right )=\left ( \begin{array}{cc}
-1 & -1 \\ -1 & ~~0 \end{array} \right )  \left ( \begin{array}{c} \kappa\\ y \end{array} \right ),
\end{equation}
where $\dot{\kappa}=d\kappa/dl$ and similarly for $\dot{y}$. Expanding $\kappa$ and $y$ in
eigenstates $A_{\pm}(l)=A_{\pm}e^{\lambda_{\pm}l}$ of the fixed-point
stability matrix above, the eigenvalues are
$\lambda_{\pm}=\pm(\sqrt{5}\mp 1)/2$. These
define the relevant and irrelevant axes in the $K_l-y_l$ plane. We
assume, following existing procedure, that the relevant scaling field
$A_+$ is the temperature axis, $A_+\approx A|\epsilon|$, where
$\epsilon=(1-T/T_c)$ is the deviation of the temperature $T$ from the bulk
transition
temperature $T_c$, and $A$ is a constant.
The rescaling law for the free energy $F$ per unit area implies
\begin{eqnarray}
Z(K_0,y_0,j, H)&= &e^{-(F_l-F_0)L^2}Z(K_l,y_l,j_l,H_l)\nonumber \\
&= &e^{-(F_l-F_0)L^2}Z\left(A|\epsilon|e^{\lambda_+l},
A_-e^{\lambda_-l}, je^l, He^{-l}\right),
\end{eqnarray}
where $Z$ is the partition function and $K_0$ and $y_0$ are the initial values of $K_l$ and $y_l$ at scale size $r_c$. 

\begin{figure}
  \centering
  \includegraphics[width=15cm]{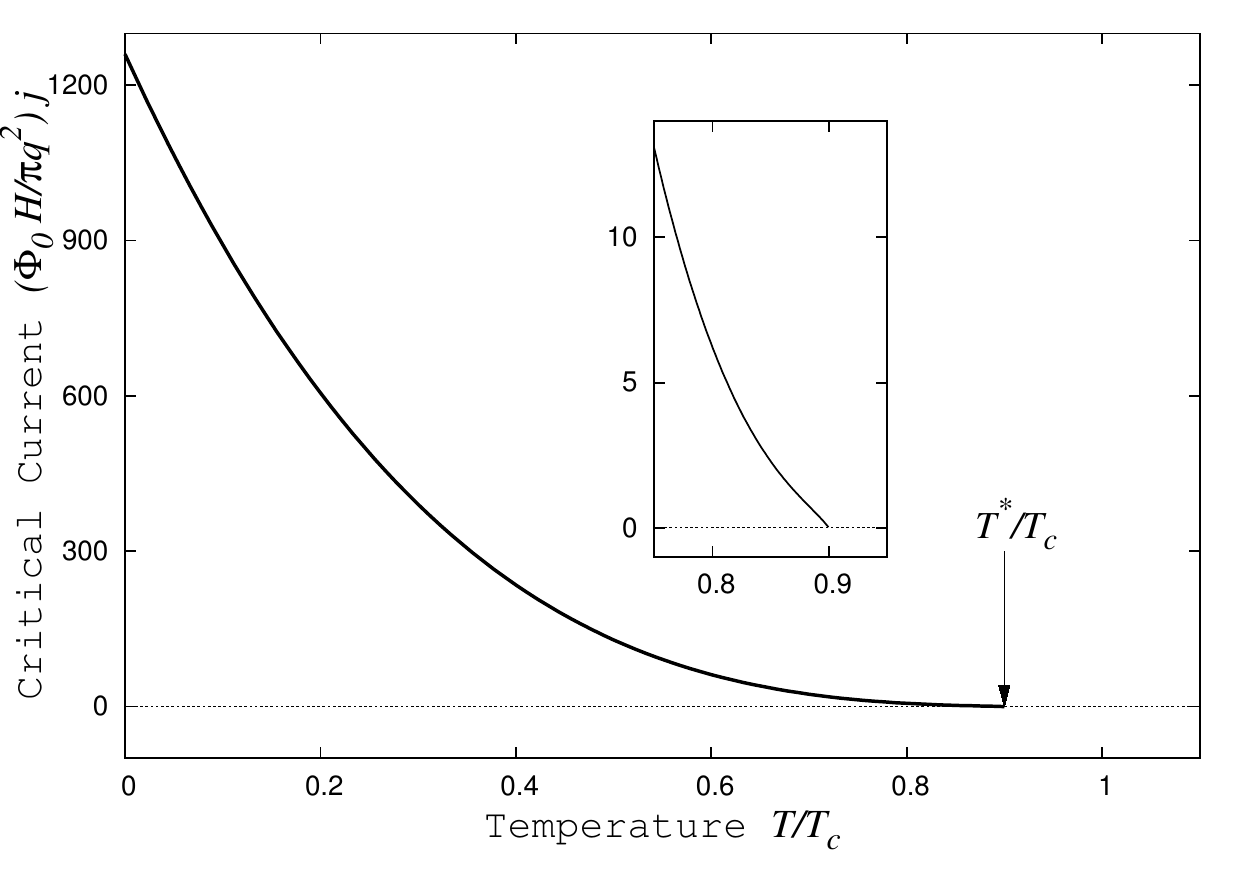}
  \caption{Plots of the dimensionless critical current $(\Phi_0 H/\pi q^2)j$ as a function of the dimensionless temperature $T/T_c$. For illustration purposes, we have taken $T^*/T_c=0.9$. The insert shows the behaviour of $j$ near $T^*$.}\label{fig:myfigure3}
\end{figure}

The scaling stops when the screening plasma of vortices completely fills the space of the channel between $x=x_0$ and $x=H$. At this point, the attraction of the vortices to the boundary at $x=H$ is completely screened. Vortices move freely from the right-hand boundary to the left-hand boundary -- there is an avalanche of phase slips. The channel becomes resistive and superconductivity is destroyed. This happens when $H_l-(x_0)_l=r_ce^l$, {\it i.e.}, when 
\begin{equation}
l=l_0=\ln(\alpha H/r_c)^{1/2},
\end{equation}
where we have written $x_0=H\alpha$ (see Eq.~(5)). To a first approximation, we ignore the scaling of the superfluid density in the argument of $\tan^{-1}$ in $x_0$. Setting $l=l_0$ in Eq.~(10), we find that the partition function is well-defined only if the coherence length diverges as
\begin{equation}
\xi(=r_ce^{l_0})=r_c|\epsilon|^{-1/\lambda_+}.
\end{equation}
Combining Eqs.~(11) and (12) gives
\begin{displaymath}
\alpha=1-\frac{2}{\pi}\tan^{-1}\frac{j\Phi_0 H}{\pi q^2}=\frac{r_c}{H}|\epsilon|^{-2/\lambda_+},
\end{displaymath}
from which the temperature-dependent critical current $j$ for the onset of phase slip events follows as
\begin{equation}
j_c=\frac{\pi q^2}{\Phi_0 H}\tan \frac{\pi}{2}\!\left (1-\frac{r_c}{H}
|\epsilon|^{-2/\lambda_+}\right )=\frac{\pi q^2}{\Phi_0 H}\cot \frac{\pi}{2}\left (\frac{1-T^*/T_c}{1-T/T_c}\right )^{\!2/\lambda_+} ~~~(T\leqslant T^*),
\end{equation}
where $T^*=T_c\left[1-(r_c/H)^{\lambda_+/2}\right]$.
Note that when $T=T^*$, $j_c=0$ as expected. 

Note that, if no current flows in the channel, nucleation of vortices will drive the system to a topological phase transition at temperature $T^*<T_c$, with the critical temperature depressed by an amount $\Delta T= (r_c/H)^{\lambda_+/2}$ that varies inversely with the width $H$ ($\gg r_c$) of the superconducting/superfluid channel.

The dependence $j_c$ on $T$ is shown in Fig.~3, and the inset shows the dependence of $j_c$ in the immediate vicinity of $T^*$. 
The result (13) shows that the temperature at which the superconductivity of the channel is destroyed in the presence of a current is smaller than $T^*$ and decreases with increasing $j$. 

In Fig.~4 we compare our result for the critical current $j_c$ per unit length Eq. (13) with the experimental data of Moon {\it et al.} \cite{Moon} on superconducting MgB$_2$ thin films of thickness $d\sim$ 440-490 nm; Moon and co-workers measured the critical current density $J_c$ per unit area ($J_c=j_c/d$) by transport methods in MgB$_2$ films on Al$_2$O$_3$, made from boron deposited at room temperature and at 750$^\circ$\,C, and on MgB$_2$ deposited on MgO. The transition temperature $T_c$ of these films are respectively ~39\,K, ~37.4\,K and ~38\,K \cite{Moon}. The comparison shows that the theory presented here describes well the experimental data in the critical region. There the critical current can be well described with the use of critical indices.
From Eqs. (12) and (13) we find that
\begin{displaymath}
j_c \sim ( T_c-T)^{2/ \lambda_+}\sim \xi^{-2},
\end{displaymath}
{\it i.e.}, the critical exponent for the phase slip catastrophe is double that of the critical exponent for the correlation length.
In Ginzburg-Landau theory, the value of the critical exponent for the correlation length is equal to $1/2$. In our case, it depends on the geometry of the film
and on the geometry of inhomogeneous structures inside the film which determines the current flow. The dependence  $ j_c\sim \xi^{-2}$ has a very simple explanation. The critical current is determined by the characteristic area of the critical fluctuations. Further our finding shows that the critical current is well described by the critical theory of phase transitions, where all are classified by critical indices.

\begin{figure}
  \centering
  \includegraphics[width=14cm]{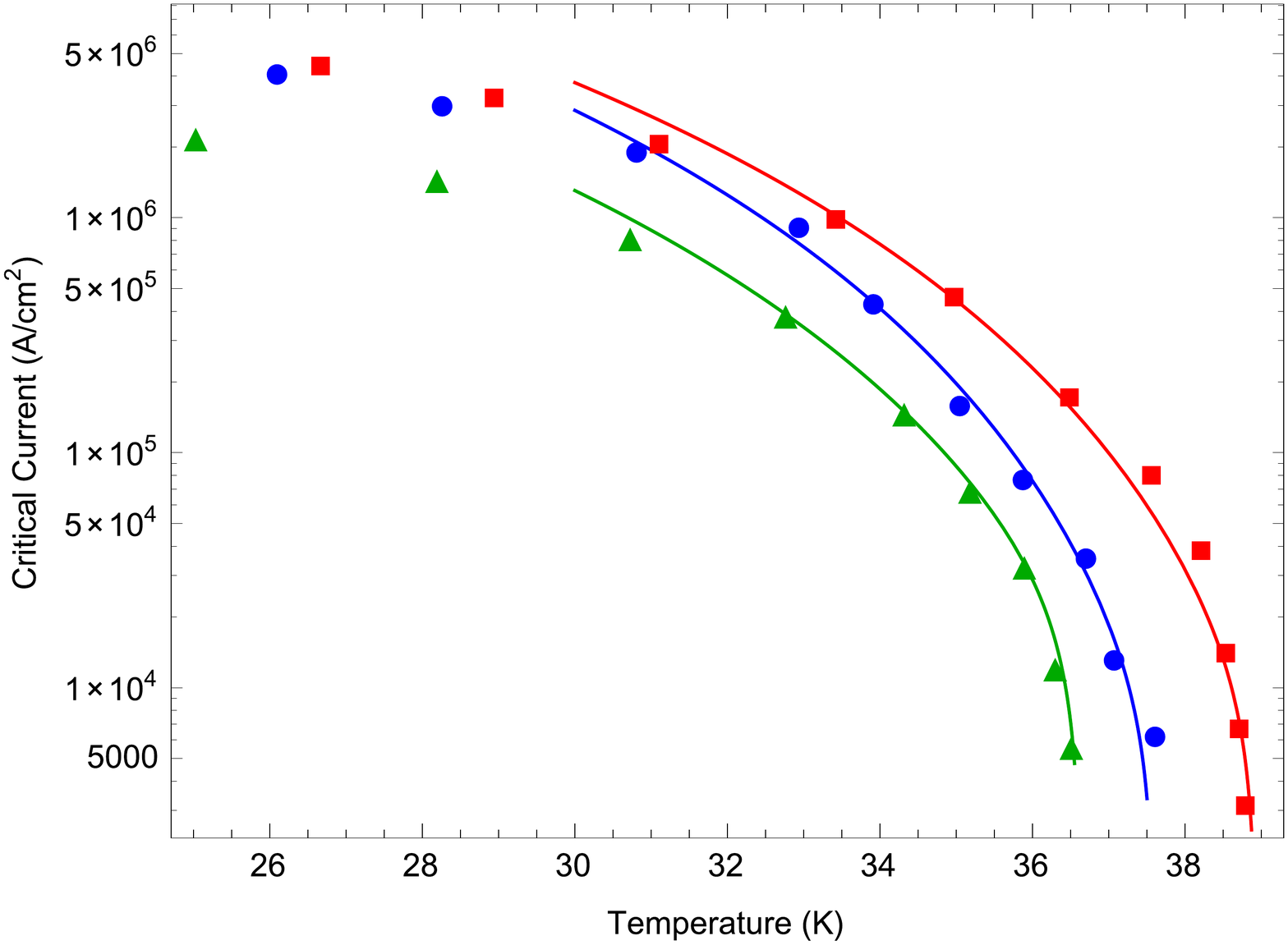}
  \caption{Critical current density $J_c$ (current per area) as a function of temperature $T$ for MgB$_2$ films deposited on Al$_2$O$_3$ (in red) and on MgO (in blue) at room temperature, and on Al$_2$O$_3$ at 750$^\circ$\,C (in green). The data points are from the experiments of Moon {\it et al.} \cite{Moon} and the continuous lines are from the analytical result Eq. (13): the films used in the experiments have thickness $d\sim$ 440-490 nm, and $J_c=j_c/d$.}\label{fig:myfigure3}
\end{figure}

We have described a mechanism of phase slip avalanches arising in superfluid or superconducting
systems.  A single vortex created in a channel or wire is strongly
attracted to the boundary. The
attraction is due to the supercurrent pattern associated with the vortex and can be well
described as a Coulomb attraction of the ``vortex charge''
to its image  anti-vortex (IA) beyond the boundary. However, the
fluctuating creation of a plasma in the space between the vortex and the
boundary screens the V-IA attraction by renormalizing the superfluid
density. This further improves the condition for the creation of more
of such single vortices and eventually there will arise so many of these that the attraction is completely screened, resulting in an avalanche of phase slip events which drives the system into a resistive state.  We have used a real-space renormalization group method to derive the
scaling laws describing such a catastrophe, and shown that this occurs at some critical current and a temperature which is lower than the
bulk transition temperature $T_c$. We have derived an exact analytical expression for the critical current as a function of film width and temperature valid at high temperatures near the critical temperature. We have analysed some of the available experimental data of critical currents in superconducting thin films \cite{Moon} and it would seem that the data show evidence of phase slip avalanches.  This phenomenon of avalanches can be further studied in small superconductors of specific shapes. For example, if the superconducting strip has a narrowing dip, then this dip can serve as a source of avalanches. Using imaging techniques, these phase slip avalanches may be visualised as those seen in the experiments by Johansen and co-workers \cite{Johansen-2002}.  We plan to propose further experiments to verify the theory we have presented here -- experiments to detect phase slip avalanches stimulated by current flow and/or temperature, and which depends on the shape and size of the small superconductor and can influence Josephson Plasma Resonance \cite{Marat-2000}. Further our result $\Delta T= (r_c/H)^{\lambda_+/2}$ for the depression of the critical in the absence of a current can be used to provide an insight on the nanostructures of granular superconductors, in particular the grain size.

We would like to stress that the mechanism described in this manuscript of the creation of the vortex avalanche catastrophe in which one vortex may generate infinitely many other vortices and anti-vortices is new and has never been discussed before. 

We are grateful to Marat Gaifullin for many useful discussions.

\end{document}